\documentclass[twocolumn]{aastex631}

\usepackage{amsmath}
\usepackage{amssymb}
\usepackage{natbib}

\newcommand{\chieff}{\chi_{\rm eff}}

\newcommand{\facc}{f_{\rm acc}}
\newcommand{\fhier}{f_{\rm hier}}
\newcommand{\fstd}{f_{\rm std}}
\newcommand{\betalow}{\beta_{\rm low}}

\providecommand{\dodoi}[1]{\doi{#1}}

\shorttitle{High-Spin BBH Subpopulation from AGN Accretion}
\shortauthors{Bartos et al.}

\begin{document}

\title{A High-Spin Subpopulation in Binary Black Hole Mergers Consistent with AGN Accretion}

\author{Imre Bartos}
\email{imrebartos@ufl.edu}
\affiliation{Department of Physics, University of Florida, Gainesville, FL 32611, USA}

\author{Zolt\'an Haiman}
\email{zoltan.haiman@ista.ac.at}
\affiliation{Institute of Science and Technology Austria (ISTA), Am Campus 1, Klosterneuburg, Austria}
\affiliation{Department of Astronomy, Columbia University, New York, NY 10027, USA}
\affiliation{Department of Physics, Columbia University, New York, NY 10027, USA}

\begin{abstract}
The formation environments of merging binary black holes remain uncertain. While hierarchical assembly in dense stellar clusters has been widely explored as an explanation for black holes exceeding the stellar-mass limit, growth through gas accretion in active galactic nucleus (AGN) disks is an alternative that has received less observational scrutiny. Here we search for an accretion-origin subpopulation using only spin magnitudes, fitting a three-component mixture model to 166 binary black hole mergers from LIGO--Virgo--KAGRA with component shapes fixed from theoretical predictions and only the mixing fractions inferred from the data. We find strong evidence ($\ln \mathcal{B}_{\mathcal{M}_1/\mathcal{M}_0} = 5.7$) that $\sim 10\%$ (90\% credible interval $[1\%, 14\%]$) of detected mergers belong to a subpopulation with primary spins clustered near $a_1 \approx 0.9$, consistent with the theoretical prediction for accretion spin-up. The hierarchical-merger prediction of $a_1 \approx 0.7$ is decisively disfavored as the location of the high-spin subpopulation ($\ln \mathcal{B}_{\mathcal{M}_1/\mathcal{M}_2} = 5.7$). Post hoc validation reveals that the accretion candidates have systematically higher masses (median $m_1 = 58\,M_\odot$) and aligned spins (median $\chi_{\rm eff} = 0.33$, vs.\ $0.04$ for standard-dominated events). The accretion subpopulation is not limited to systems above the pair-instability mass gap: GW190517 ($m_1 \approx 39\,M_\odot$) is among the top candidates, demonstrating that accretion spin-up operates across a range of masses. GW190521, previously interpreted as a hierarchical merger, shows comparable support for
an accretion origin. These results provide the first population-level observational evidence for an accretion-origin subpopulation in black hole mergers.
\end{abstract}


\section{Introduction}
\label{sec:intro}

Gravitational-wave observations by the LIGO, Virgo, and KAGRA 
observatories have revealed a population of merging black holes 
with masses that challenge standard stellar-evolution expectations
\citep{2015CQGra..32g4001L,VIRGO:2014yos,KAGRA:2018plz}.
Stellar evolution models predict an upper mass limit for black 
holes formed via direct collapse at roughly 
$\sim$50\,$M_\odot$, due to pulsational and pair-instability 
supernovae 
\citep{Heger2002,Woosley2017,Farmer2019}.
Despite this expectation, several gravitational-wave events 
contain black holes with component masses at or well above 
this limit.
Notable examples include GW190521, with component masses 
$m_1 = 85^{+21}_{-14}\,M_\odot$ and 
$m_2 = 66^{+17}_{-18}\,M_\odot$, and GW231123, the most 
massive binary black hole merger detected to date, with 
$m_1 = 137^{+23}_{-18}\,M_\odot$ and 
$m_2 = 101^{+22}_{-51}\,M_\odot$ 
\citep{Abbott2020GW190521,Abbott2025GW231123}.
These observations indicate that at least a subset of merging 
black holes cannot be explained by isolated stellar collapse alone.

One way to reach such large black hole masses is through 
successive mergers of smaller black holes.
In this picture, black holes formed via stellar collapse merge, 
and the merger remnants subsequently participate in additional 
mergers \citep{Fishbach2017,GerosaBerti2017}.
This hierarchical growth channel has been explored extensively 
in dense stellar environments such as globular clusters 
\citep{Rodriguez2019}, as well as in nuclear star clusters 
and active galactic nuclei 
\citep{Yang2019,Tagawa2021hierarchical}.
Hierarchical mergers can naturally populate the mass range 
above the pair-instability limit and have therefore been widely 
invoked to explain the most massive gravitational-wave events.

An alternative possibility is that black holes experience 
substantial mass growth through gas accretion prior to merger.
Accretion can increase black hole masses beyond their birth 
values without requiring repeated mergers \citep{Levin2007}.
Such growth may operate efficiently in gas-rich environments, 
most notably in active galactic nuclei 
\citep{Bartos2017,Stone2017,Yi2018} or in the dense inner regions of protogalaxies~\citep{2020ApJ...903L..21S}.
Accretion can also modify the properties of merging binaries 
during their inspiral, including their orbital evolution and 
spin magnitudes \citep{Tagawa2020,Yang2020}. Population-level phase-space analyses have also identified individual events whose parameters are consistent with accretion-grown progenitors \citep{2025PhRvD.112b3531A}.

The existence of black holes above the pair-instability limit 
does not, by itself, uniquely identify the underlying growth 
mechanism.
However, spin magnitudes can provide complementary constraints.
In hierarchical growth, merger remnants characteristically 
acquire dimensionless spins of order $\sim$0.7, and repeated 
mergers do not trivially drive spins to more extreme values 
\citep{Pretorius2005,Tagawa2021hierarchical}.
Moreover, hierarchical mergers often involve a merger remnant 
paired with a lower-mass, first-generation companion, so 
mass ratios closer to $\sim$0.5 are expected to be common 
\citep{GerosaBerti2017,Tagawa2021hierarchical}.
In contrast, sustained accretion can efficiently spin up black holes to high values. Thin-disk accretion drives spins toward the Thorne limit of 
$a \approx 0.998$ \citep{Thorne1974}, modestly reduced by 
magnetic torques in the inner accretion flow \citep{Gammie2004}. X-ray observations of supermassive black holes find a significant population with spins $a > 0.9$, consistent with 
growth through coherent accretion \citep{Reynolds2021}.
Accretion in binaries can also modify mass ratios through 
preferential growth of the lower-mass secondary, which tends 
to equalize the component masses over time 
\citep{Young2015,Duffell2020,Siwek2023}.
These considerations suggest that accretion-assisted binaries 
may preferentially populate high-spin, near-equal-mass systems, 
in contrast to hierarchical growth.

Several recent population analyses have begun to probe the 
relationship between mass ratio and spin in the 
gravitational-wave data.
Studies have reported evidence that the effective aligned spin 
parameter $\chi_{\rm eff}$ depends on mass ratio, with 
unequal-mass systems exhibiting systematically larger 
$\chi_{\rm eff}$ 
\citep{Callister2021,Adamcewicz2022,McKernan2022}.
\citet{VijaykumarFishbach2026} showed that a hierarchical 
subpopulation concentrated around $q \sim 0.5$ with spins 
near $\sim$0.7 can generate an apparent population-level 
$q$--$\chi_{\rm eff}$ relation, demonstrating that 
hierarchical growth alone can reproduce such a trend without 
invoking additional channels.
However, accretion-driven growth occupies a distinct region 
of parameter space: spin magnitudes are expected to be 
systematically higher than the hierarchical prediction, while 
mass ratios may evolve toward unity through preferential 
growth of the secondary.
The observed population-level $q$--$\chieff$ structure need not arise from a direct correlation within any single channel, 
but can instead reflect the differing properties of subpopulations contributing to the overall mixture.

The structure of 
the high-mass population provides an additional diagnostic.
At the high-mass end, systems consistent with high spins 
appear more likely to occupy near-equal mass ratios than 
lower-spin systems in the overall population 
\citep{RayKalogera2025}.
If high masses were produced predominantly through asymmetric 
hierarchical mergers, one would expect many such systems to 
have mass ratios significantly below unity.
The presence of high-spin, near-equal-mass systems is 
therefore less straightforward to reconcile with purely 
hierarchical growth.
In contrast, accretion-driven evolution naturally links high 
spins with more equal mass ratios, since sustained accretion 
can both spin up black holes and preferentially increase the 
mass of the secondary.

Population-level analyses have also reported that inferred 
spin magnitudes increase with primary mass, with higher-mass 
systems showing stronger evidence for substantial spins 
\citep{Pierra2024,GWTC4pop}.
This trend is difficult to attribute to birth properties alone 
and instead suggests that processes which modify black holes 
after formation become increasingly important toward the 
high-mass end. 

Beyond correlations between spin and other parameters, direct modeling of the joint $(a_1, a_2)$ distribution has revealed a distinct high-spin subpopulation. \citet{2026ApJ...996...71H} and \citet{2025ApJ...994..261A} identify a component comprising $\sim$15--20\% of binaries in which both black holes have large spins.

Despite these findings, the role of accretion-driven growth in shaping the observed population has not been systematically assessed. Hierarchical mergers have been incorporated into statistical models of the gravitational-wave catalog and shown to reproduce several observed trends \citep{VijaykumarFishbach2026}, including most recently the high-spin, high-mass component in GWTC-4 \citep{2026arXiv260505563X}.  Accretion, which can simultaneously account for mass growth above the pair-instability limit and the high spins expected from sustained disk accretion, has not been incorporated as a competing population component within a unified statistical inference framework. As a result, the extent to which accretion contributes to the 
high-mass, high-spin population remains unclear.

In this work, we search for an accretion-origin subpopulation at the population level. Because the theoretical predictions for both hierarchical and accretion channels carry substantial uncertainties in their detailed mass and spin distributions, we adopt a deliberately minimalist approach: we construct a three-component mixture model using only the most robust spin predictions from each channel, fix the component shapes from theory, and infer only the mixing fractions from the data. By fitting spin magnitudes alone and validating against masses, mass ratios, and spin tilts post hoc, we obtain a clean test for the existence of an accretion subpopulation that is independent of mass modeling assumptions.

\section{Method}
\label{sec:method}

\subsection{Population Model}
\label{sec:model}

We model the population distribution of binary black hole 
component spin magnitudes as a mixture of three components, 
each representing a formation channel:
\begin{multline}
p(a_1, a_2 \mid \boldsymbol{\theta}) = 
\fstd\, p_{\rm std}(a_1, a_2) 
+ \fhier\, p_{\rm hier}(a_1, a_2) \\
+ \facc\, p_{\rm acc}(a_1, a_2),
\label{eq:mixture}
\end{multline}
where $\fstd = 1 - \fhier - \facc$, and $a_1$ and $a_2$ 
denote the dimensionless spin magnitudes of the primary 
(more massive) and secondary black holes, respectively.

The \textit{standard} component represents all formation 
channels that produce low natal spins \citep{Fuller2019,Belczynski2020},
including isolated 
binary evolution, common envelope evolution, and dynamical 
capture.
For this component, both spins are drawn from a Beta distribution peaked at zero:
$a_1, a_2 \sim \mathrm{Beta}(1, \betalow)$,
where $\betalow$ is a free parameter controlling how 
steeply the distribution falls away from zero.

The \textit{hierarchical} component represents 
second-generation--first-generation (2g--1g) mergers, the 
dominant hierarchical channel.
For this component, the primary spin follows a truncated normal distribution 
centered on the numerical relativity prediction for merger 
remnant spins:
$a_1 \sim \mathrm{TN}(0.7, 0.1)$ on $[0, 1]$.
The value $a_1 \approx 0.7$ corresponds to the remnant spin 
from equal-mass, non-spinning progenitors; unequal mass ratios 
produce somewhat lower remnant spins, but the $\sigma = 0.1$ 
width accommodates this variation 
\citep{GerosaBerti2017,PhysRevD.81.084023}.
The secondary, being a first-generation black hole, is assumed to have 
low spin drawn from the same distribution as the standard 
channel: $a_2 \sim \mathrm{Beta}(1, \betalow)$.

The \textit{accretion} component represents black holes 
that have been spun up through sustained gas accretion, 
as expected in AGN disk environments.
For this component, both spins follow a truncated normal distribution near the 
theoretical spin-up equilibrium:
$a_1, a_2 \sim \mathrm{TN}(0.9, 0.1)$ on $[0, 1]$.
The fiducial peak value of $\mu = 0.9$ is chosen as a round 
number within the range of observed supermassive black hole 
spins and below the theoretical Thorne limit.
The qualitative results are robust to this choice  
(Section~\ref{sec:robustness}).

The model has three free parameters: the hierarchical and 
accretion mixing fractions $\fhier$ and $\facc$, and the 
low-spin shape parameter $\betalow$.

The peak locations and qualitative structure of each component 
are motivated by theoretical predictions, while the specific 
parametric forms (Beta and truncated normal distributions) are arbitrary and are
chosen for simplicity. We verify the sensitivity of our results 
to these choices through extensive robustness checks 
(Section~\ref{sec:robustness}), including variation of the peak 
widths and the functional form of the standard component.

\subsection{Hierarchical Bayesian Inference}
\label{sec:inference}

We employ hierarchical Bayesian inference to constrain the 
population parameters using individual event posterior samples 
\citep{Thrane2019,Mandel2019}.
The source-parameter vector for event $i$ is 
$\theta = (m_1, q, z, a_1, a_2, \ldots)$, and posterior samples 
are drawn under the LVK parameter estimation prior 
$\pi_{\rm PE}(\theta)$.
Since we infer only the spin-sector population 
$p_{\rm spin}(a_1, a_2 \mid \boldsymbol{\theta})$ rather than 
performing a joint mass-spin-redshift fit, we condition on a 
fixed reference population over $(m_1, q, z)$ taken from the 
GWTC-4 best-fit binary black hole population analysis 
\citep{GWTC4pop}:
\begin{equation}
p_{\rm ref}(m_1, q, z) 
  = p_{\rm ref}^{m_1}(m_1)\,
    p_{\rm ref}^{q}(q \mid m_1)\,
    p_{\rm ref}^{z}(z),
\label{eq:pref}
\end{equation}
with $p_{\rm ref}^{m_1}$ a Broken Power Law $+$ 2 Peaks model, 
$p_{\rm ref}^{q} \propto q^{\beta_q}$ with $\beta_q = 1.2$, and 
$p_{\rm ref}^{z}(z) \propto (1+z)^{\kappa-1}\, dV_c/dz$ with 
$\kappa = 3.2$. All hyperparameters are fixed at the median of the GWTC-4 BBH 
hyperposterior \citep{GWTC4pop}.
Robustness to this choice is examined in 
Section~\ref{sec:robustness}.

Under this factorization the population over the full 
source-parameter vector is 
$p(\theta \mid \boldsymbol{\theta}) 
  = p_{\rm ref}(m_1, q, z)\, 
    p_{\rm spin}(a_1, a_2 \mid \boldsymbol{\theta})$, and the 
per-event marginal likelihood becomes
\begin{equation}
\mathcal{L}_i(\boldsymbol{\theta}) 
  \propto \frac{1}{n_i} \sum_{j=1}^{n_i}
    w_{ij}\, 
    p_{\rm spin}\!\left(a_1^{(i,j)}, a_2^{(i,j)} 
      \mid \boldsymbol{\theta}\right),
\label{eq:event_lhood}
\end{equation}
where the importance-sampling weights account for the mismatch 
between the reference population and the PE prior in the 
$(m_1, q, z)$ sector:
\begin{equation}
w_{ij} 
  = \frac{p_{\rm ref}\!\left(m_1^{(i,j)}, q^{(i,j)}, z^{(i,j)}\right)}
         {\pi_{\rm PE}^{m,q,z}\!\left(m_1^{(i,j)}, q^{(i,j)}, z^{(i,j)}\right)}.
\label{eq:weights}
\end{equation}
The LVK parameter estimation analyses adopt a prior that is 
uniform in the detector-frame component masses and uniform in 
source-frame comoving volume per source-frame time for the 
distance, which in $(m_1^{\rm source}, q, z)$ coordinates 
corresponds to 
$\pi_{\rm PE}^{m,q,z}(m_1, q, z) \propto m_1\,(1+z)\,dV_c/dz$.
The PE prior on spin magnitudes is uniform on $[0, 0.99]$ for 
both components with isotropic spin orientations, identical to 
the spin sector of $p_{\rm ref}$, so spin contributes only a 
multiplicative constant that cancels in the per-event 
normalization.

The reliability of the importance-sampling reweighting is 
monitored per-event via the effective sample size,
\begin{equation}
\mathrm{ESS}_i 
  = \frac{\left(\sum_j w_{ij}\right)^2}{\sum_j w_{ij}^2}.
\label{eq:ess_event}
\end{equation}
Events for which $\mathrm{ESS}_i / n_i$ falls below $10\%$ are 
flagged as having reweighted likelihoods dominated by a small 
fraction of their posterior samples; for these events the 
per-event memberships in Section~\ref{sec:candidates} are 
reported as indicative rather than as point estimates.
Under the fiducial reference, $7$ of the $166$ events fall 
below this threshold, none of which lie in the top accretion 
candidates.

We adopt a uniform prior on $(\fhier, \facc)$ subject to 
$\fhier \geq 0$, $\facc \geq 0$, $\fhier + \facc \leq 1$, and a 
log-uniform prior on $\betalow \in [1.5, 40]$.
The posterior is computed on a $50^3$ grid in 
$(\fhier, \facc, \betalow)$ with $\betalow$ logarithmically 
spaced, providing exact marginalization without 
sampler-convergence concerns. Per-grid-point hyperparameters 
are retained in the posterior only when the selection-function 
effective sample size satisfies the criterion described in 
Section~\ref{sec:selection}. Posterior summaries are reported as the maximum a posteriori (MAP) estimate and $90\%$ credible interval (computed from the marginalized cumulative distribution), with the caveat that the grid spacing limits the MAP precision (e.g.\ $\Delta \facc \approx 0.02$).

Model comparison is performed via Bayes factors computed from 
grid-integrated evidences for nested models:
$\mathcal{M}_0$ (standard only), 
$\mathcal{M}_1$ (standard $+$ accretion), 
$\mathcal{M}_2$ (standard $+$ hierarchical), and 
$\mathcal{M}_3$ (all three components).
Each evidence integral incorporates the catalog-level selection 
correction described in Section~\ref{sec:selection}.

Per-event membership probabilities are computed by averaging 
over the full three-dimensional posterior grid, weighted by the 
importance-sampled per-event likelihood, rather than evaluating 
at a single point estimate. For event $i$, the probability of 
belonging to component $k$ is
\begin{equation}
p_k^{(i)} 
  = \int p_k^{(i)}(\boldsymbol{\theta})\, 
    p(\boldsymbol{\theta} \mid \mathbf{d})\, 
    d\boldsymbol{\theta},
\label{eq:membership}
\end{equation}
where $p_k^{(i)}(\boldsymbol{\theta})$ is the membership 
probability at fixed population parameters and 
$p(\boldsymbol{\theta} \mid \mathbf{d})$ is the population 
posterior. The same importance-sampling weights $w_{ij}$ enter 
the per-event sum at fixed $\boldsymbol{\theta}$.

\subsection{Data and Waveform Selection}
\label{sec:data}

We analyze 166 binary black hole events from the GWTC-2.1 
\citep{GWTC21}, GWTC-3 \citep{GWTC3}, and GWTC-4.0 
\citep{GWTC4} catalogs.
Where available, we use posterior samples obtained with the 
NRSur7dq4 numerical relativity surrogate model 
\citep{NRSur7dq4}, which is calibrated directly to numerical 
relativity simulations within its domain of validity.
NRSur7dq4 samples are available for 90 events: 43 from the 
official GWTC-4.0 release, 46 from the NRSurCat-1 reanalysis 
of O1--O3 events \citep{NRSurCat1}, and GW190521 from its 
dedicated discovery analysis \citep{GW190521}.
For the remaining 76 events, which fall outside the NRSur7dq4 
calibration region, we use the mixed-approximant posterior 
samples provided in the standard catalog releases.
We examine the sensitivity of our results to waveform choice 
in Section~\ref{sec:robustness}.

\subsection{Selection Effects}
\label{sec:selection}

We compute the detection efficiency $\alpha(\boldsymbol{\theta})$ via importance-sampling reweighting of the LVK O3$+$O4a injection set \citep{GWTC4pop}, applying a network signal-to-noise ratio threshold of $\mathrm{SNR} > 10$ following the standard LVK convention. The selection-function effective sample size $\mathrm{ESS}_\alpha$ is monitored at each grid point, and hyperparameter values with $\mathrm{ESS}_\alpha < 4\,N_{\rm obs}$ are excluded. Across the spin-model parameter ranges of Section~\ref{sec:robustness} 
the detection efficiency varies by approximately $25\%$; the hyperparameter posteriors incorporate this via the catalog-level form of the population likelihood under detection selection \citep{Mandel2019},
\begin{equation}
\ln \mathcal{L}_{\rm pop}(\boldsymbol{\theta})
  = \sum_{i=1}^{N_{\rm obs}} \ln \mathcal{L}_i(\boldsymbol{\theta})
    \;-\; N_{\rm obs}\, \ln \alpha(\boldsymbol{\theta}).
\label{eq:pop_lhood}
\end{equation}

\section{Results}
\label{sec:results}

\subsection{Spin-Only Analysis}
\label{sec:spinonly}

\begin{figure}
    \centering
    \includegraphics[width=\columnwidth]{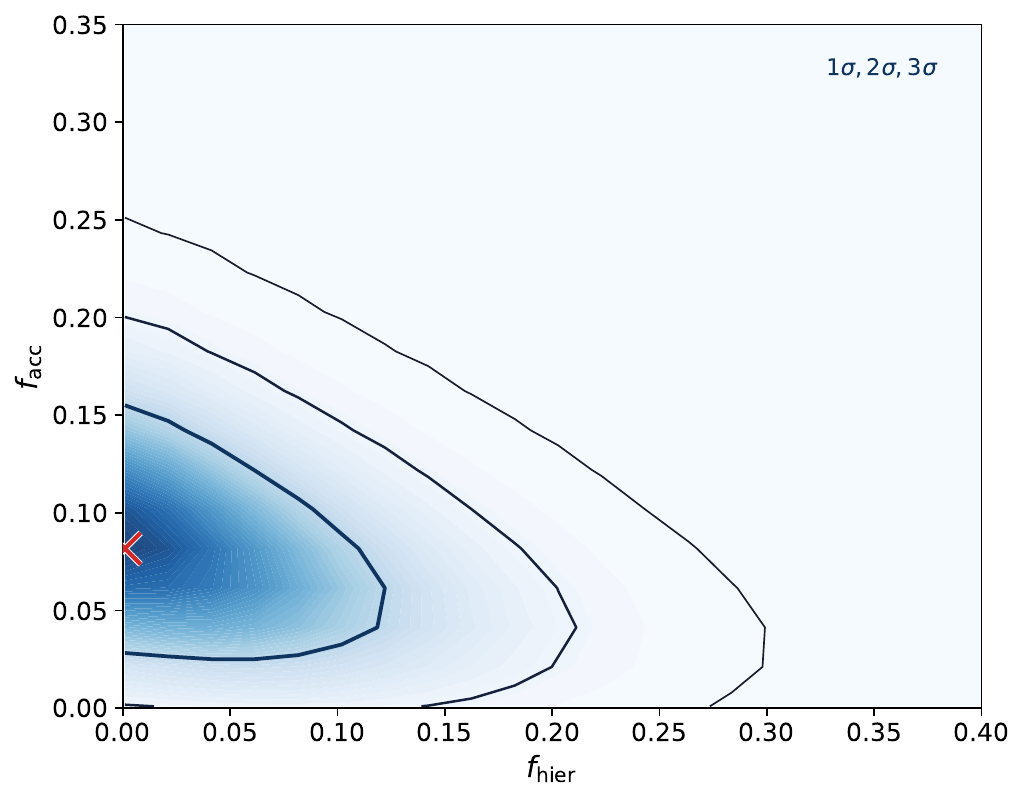}
    \caption{{\bf Joint posterior on the hierarchical and accretion mixing 
fractions from the spin-only analysis}, marginalized over $\betalow$. Contours show $1\sigma$, $2\sigma$, and $3\sigma$ credible regions. The posterior is concentrated along $\fhier \approx 0$, indicating no significant detection of the hierarchical channel from spins alone, while $\facc$ is clearly excluded from zero (MAP 8.2\%, 90\% CI [1.4\%, 13.7\%]).}
    \label{fig:detection}
\end{figure}

The spin-only analysis yields strong evidence for an accretion
subpopulation.
The marginal posterior on $\facc$ excludes zero, with a maximum
a posteriori (MAP) estimate of $\facc = 8.2\%$ and a 90\%
credible interval $[1.4\%, 13.7\%]$
(Figure~\ref{fig:detection}).
The MAP estimates of the other model parameters are $\fhier = 0$
and $\betalow = 5.0$; the marginal posterior on $\fhier$ peaks
at the lower grid boundary and is consistent with zero.
The Bayes factor for the standard $+$ accretion model over
standard-only is $\ln \mathcal{B} = 5.7$, constituting strong
evidence.

The hierarchical component is not detected in the spin-only 
analysis ($\ln \mathcal{B} \approx 0$ for standard + 
hierarchical vs.\ standard-only), and adding a hierarchical 
component to the standard + accretion model is disfavored 
($\ln \mathcal{B} = -2.3$).

To distinguish accretion- from hierarchical-channel origins of the high-spin subpopulation, we compare the standard+accretion model $\mathcal{M}_1$ (peak at $a_1 = 0.9$) to the standard+hierarchical model $\mathcal{M}_2$ (peak at $a_1 = 0.7$). The data decisively favor the accretion-channel location 
($\ln \mathcal{B}_{\mathcal{M}_1/\mathcal{M}_2} = 5.7$), consistent with theoretical predictions for accretion-driven spin-up in AGN disks \citep{Tagawa2020,Yang2020} and with the typical spins of accreting supermassive black holes \citep{Reynolds2021}.

\subsection{Properties of the Accretion Subpopulation}
\label{sec:properties}

\begin{figure}[t]
\centering
\includegraphics[width=\columnwidth]{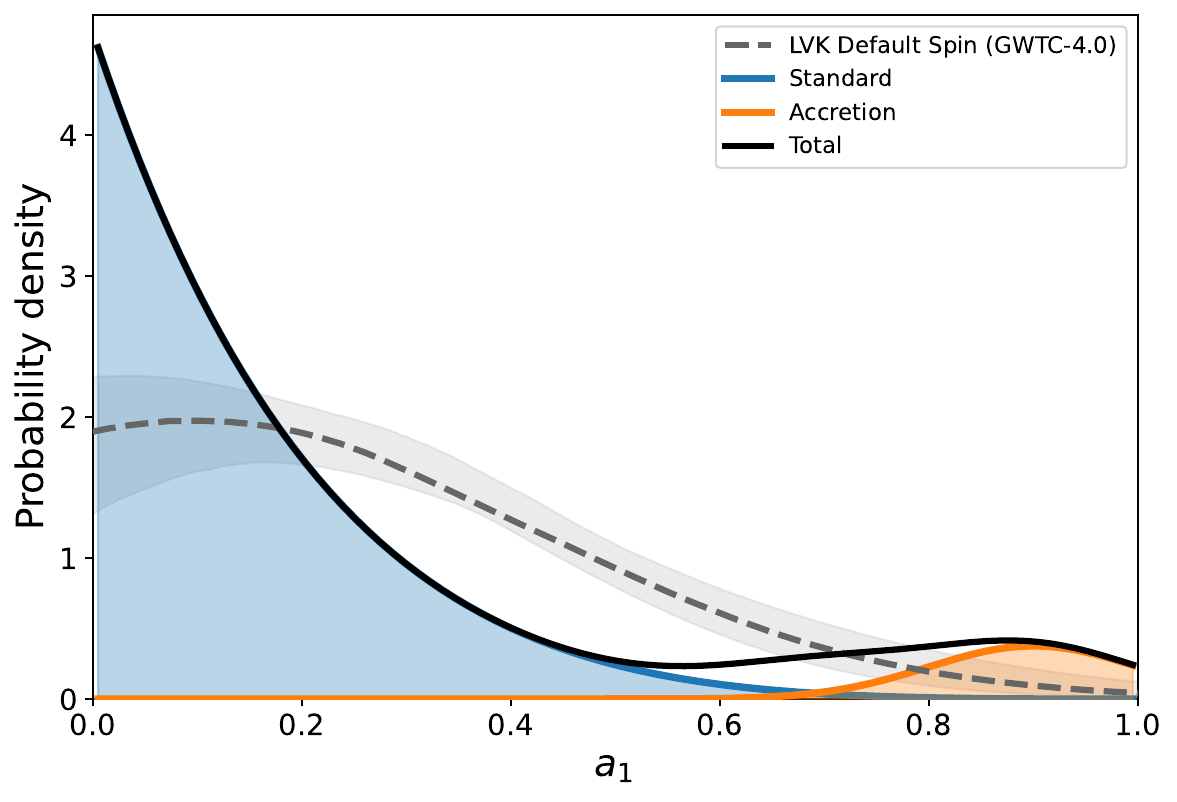}
\caption{Posterior predictive spin magnitude distributions, marginalized over parameter uncertainty. The standard component (blue, $\sim$90\%) peaks at zero and decays monotonically. The accretion $a_1$ distribution (orange, $\sim$10\%) peaks near $a_1 = 0.9$, well separated from the bulk population. The accretion $a_2$ distribution (green band) is weakly constrained by current data. The black line shows the total $a_1$ mixture. All curves are weighted by their population fractions so that the integrated areas reflect the relative prevalence of each component. The gray 
dashed line and band show the median and 90\% credible interval of the 
spin magnitude distribution from the GWTC-4.0 population analysis 
\citep{GWTC4pop}; our mixture model provides a physical decomposition 
of the same structure.}
\label{fig:intrinsic}
\end{figure}

The inferred posterior-predictive primary spin distribution 
under the canonical mixture model is shown in 
Figure~\ref{fig:intrinsic}.
The standard component peaks at zero with 
$\betalow \approx 5.0$, corresponding to a mean spin 
magnitude of $\sim$0.17.
The accretion component appears as a peak near 
$a_1 \approx 0.9$, well separated from the bulk population.

\begin{figure}[t]
\centering
\includegraphics[width=\columnwidth,trim=0 0 470 0,clip]{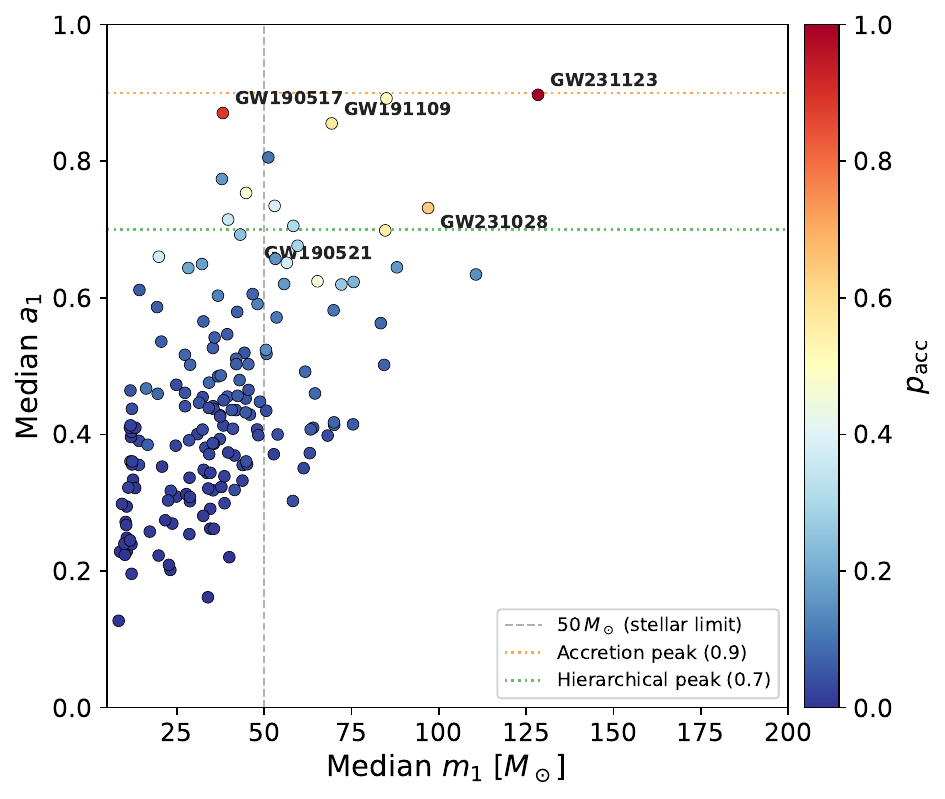}
\caption{Median posterior primary mass versus median posterior primary spin magnitude for all 166 events, colored by accretion membership probability $p_{\rm acc}$. Accretion candidates cluster at high spin, with several above the pair-instability stellar limit ($\sim$50\,$M_\odot$, dashed). Horizontal lines indicate the accretion (0.9) and hierarchical (0.7) spin predictions. Key accretion candidates are labeled. GW190521 appears near the hierarchical prediction at its median $a_1$, but both its primary and secondary spin posteriors are broad with significant support at high values.}\label{fig:validation}
\end{figure}

As a post hoc validation, we examine properties of the accretion candidates that were not used in the fit. These properties are consistent with accretion from a prograde coplanar circumbinary disk, as expected in AGN environments (Figure~\ref{fig:validation}). The accretion candidates ($p_{\rm acc} > 0.3$) have 
systematically higher masses (median $m_1 = 58\,M_\odot$ vs.\ 
$36\,M_\odot$ for the catalog) and median $\chieff = 0.33$, 
compared to 0.04 for the standard-dominated population, 
indicating preferentially aligned spins as predicted by 
accretion from a coplanar circumbinary disk 
\citep{Bogdanovic2007,King+2005}.

\subsection{Individual Candidates}
\label{sec:candidates}

\begin{figure*}[t]
\centering
\includegraphics[width=\textwidth]{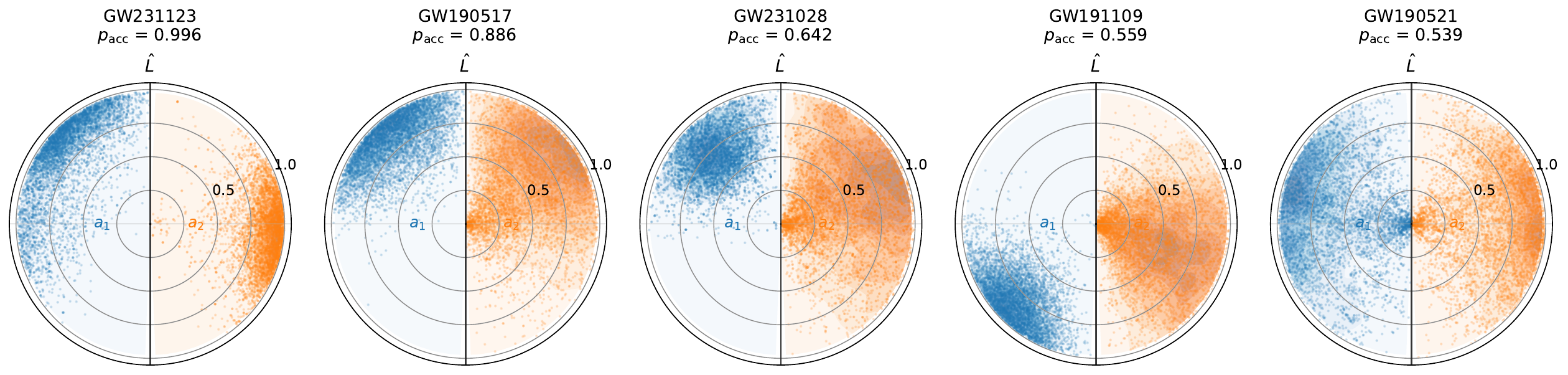}
\caption{Spin magnitude and tilt posterior distributions for the top five accretion candidates, displayed in the standard LVK polar representation. Radial distance indicates spin magnitude ($a = 0$ at center, $a = 1$ at rim); polar angle indicates tilt relative to the orbital angular momentum $\hat{L}$ ($0^\circ$ = aligned at top, $180^\circ$ = anti-aligned at bottom). Left half-disk: primary spin $a_1$ (blue); right half-disk: secondary spin $a_2$ (orange). Accretion is expected to produce high spins ($a \gtrsim 0.9$) with preferentially (anti-)aligned orientations (although other orientations may be possible; see discussion in \citealt{2026ApJ...996L..44B}), whereas hierarchical mergers predict spins near $a \approx 0.7$ with broader tilt distributions. Large aligned/antialigned spins, consistent with accretion, are favored in most cases.}\label{fig:spindisks}
\end{figure*}

Figure~\ref{fig:spindisks} shows the spin magnitude and tilt distributions for the top five accretion candidates. GW231123, the most massive BBH merger detected to date, is the most confident accretion candidate, with 
$p_{\rm acc} > 0.97$ across all analysis variants.
Both spins are high ($a_1 = 0.90$, $a_2 = 0.91$ with 
NRSur7dq4). 

GW190517, with $m_1 \approx 39\,M_\odot$ \citep{GWTC21}, is notable as an accretion candidate below the pair-instability mass gap. Its high primary spin ($a_1 = 0.87$) is characteristic of the accretion channel, demonstrating that accretion spin-up can occur without requiring growth to extreme masses.

GW231028, with $m_1 \approx 97\,M_\odot$ and $q = 0.53$, 
receives $p_{\rm acc} = 0.64$. Its primary spin of 
$a_1 = 0.73$ is moderately high, and its primary mass above 
the pair-instability gap is consistent with growth through 
gas accretion in an AGN disk.

GW191109 ($p_{\rm acc} = 0.56$, $m_1 \approx 65\,M_\odot$) sits near the boundary of the pair-instability mass gap.

GW190521 \citep{Abbott2020GW190521}, widely interpreted as a hierarchical merger based on its location in the pair-instability mass gap, receives $p_{\rm acc} = 0.54$. Although the median spins ($a_1 = 0.70$, $a_2 = 0.70$) nominally coincide with the hierarchical prediction, both posteriors are broad with significant support at higher values, and its mass ratio is near unity ($q = 0.79$), all consistent with an accretion origin.

\section{Robustness}
\label{sec:robustness}

We perform extensive sensitivity checks to verify that the accretion detection is not driven by specific modeling choices (Table~\ref{tab:robustness}).

\textit{Accretion peak width.} Varying the assumed width from 
$\sigma_{\rm acc} = 0.05$ to 0.30, the accretion fraction ranges from 8\% to 12\% and $\ln\mathcal{B}(\mathcal{M}_1/\mathcal{M}_0)$ from 5.5 to 6.3, all within the canonical $90\%$ credible interval.

\textit{Flexible standard component.} Replacing the standard 
Beta(1, $\betalow$) with a two-parameter Beta($\alpha$, $\beta$) distribution gives a slightly elevated accretion fraction ($\facc = 10.3\%$, $\ln \mathcal{B} = 7.6$). The data weakly prefer $\alpha \approx 2.7$ (90\% CI $[1.2, 4.7]$), indicating a non-monotonic low-spin distribution with a small non-zero mode.

\textit{Waveform systematics.} Using mixed-approximant samples for all events instead of the hybrid NRSurrogate selection gives $\facc = 8.2\%$ and $\ln\mathcal{B} = 5.0$. Using SEOBNRv5/v4PHM as the preferred waveform where NRSurrogate is unavailable gives $\facc = 8.2\%$ and $\ln\mathcal{B} = 4.3$. Both are consistent with the main analysis. The hybrid NRSurrogate analysis, which uses the most accurate available waveform for each event, is adopted as the primary result.

\begin{deluxetable}{lcc}
\tablecaption{Robustness Checks\label{tab:robustness}}
\tablehead{
  \colhead{Variant} & 
  \colhead{$\facc$} & 
  \colhead{$\ln \mathcal{B}_{\mathcal{M}_1/\mathcal{M}_0}$}
}
\startdata
Fiducial (hybrid NRSur) & 8.2\% & 5.7 \\
$\sigma_{\rm acc} = 0.05$ & 8.2\% & 6.3 \\
$\sigma_{\rm acc} = 0.15$ & 10.2\% & 5.7 \\
$\sigma_{\rm acc} = 0.30$ & 12.2\% & 5.5 \\
$\sigma_{\rm hier} = 0.30$ & 8.2\% & 5.7 \\
Flex.\ standard Beta($\alpha$, $\beta$) & 10.3\% & 7.6 \\
Mixed-only waveforms & 8.2\% & 5.0 \\
SEOB-priority waveforms & 8.2\% & 4.3 \\
\enddata
\tablecomments{$\facc$ is the MAP estimate. $\ln\mathcal{B} = \ln\mathcal{B}(\mathcal{M}_1/\mathcal{M}_0)$ is the log Bayes factor for standard + accretion vs.\ standard only. All configurations use 166 events and apply the catalog-level selection correction described in Section~\ref{sec:selection}.}
\end{deluxetable}

\textit{Hierarchical component width.} Broadening the hierarchical component to $\sigma_{\rm hier} = 0.3$, well beyond the physically expected spread for 2g--1g remnant spins \citep{PhysRevD.81.084023}, does not diminish the accretion signal ($\facc = 8.2\%$, $\facc = 0$ excluded at 90\%).

\section{Summary and Conclusions}
\label{sec:conclusions}

We searched for an accretion-origin subpopulation in the 
spin magnitudes of 166 binary black hole mergers using a 
three-component mixture model. The data provide strong evidence 
($\ln \mathcal{B} = 5.7$) that $\sim10\%$ (90\% credible 
interval $[1\%, 14\%]$) of mergers belong to a subpopulation 
with primary spins clustered near $a_1 \approx 0.9$, consistent 
with theoretical predictions for accretion spin-up in AGN disks. 
The hierarchical-merger prediction of $a_1 \approx 0.7$ is 
decisively disfavored as the location of the high-spin 
subpopulation ($\ln \mathcal{B}_{\mathcal{M}_1/\mathcal{M}_2} = 5.7$).

The inferred accretion fraction of $\sim$10\% for detected events is consistent with theoretical estimates of the AGN contribution to BBH mergers, which span roughly 1--25\% depending on assumptions about disk properties and migration rates \citep{Bartos2017,Yang2020rate,McKernan2020,Tagawa2020,2021ApJ...920L..42G,Ford2022,2024arXiv241212086R}. Published AGN-channel rate estimates corresponding to this range are
of order a few to $\sim$25\,Gpc$^{-3}$\,yr$^{-1}$
\citep{Bartos2017,McKernan+2018,Tagawa2020,
Yang2020rate,2021ApJ...920L..42G}.

Our mixture model provides a physical interpretation of the spin structure reported in GWTC-4.0 population analyses \citep{GWTC4pop,Heinzel2024}, which find a low-spin peak with a high-spin tail containing $\sim$20\% of binaries. In our decomposition, the low-spin peak corresponds to the standard channel ($\sim$90\%) and the high-spin tail to the accretion subpopulation ($\sim$10\%). The difference in the inferred spinning fraction likely reflects the broader definition used in those analyses, which do not distinguish between moderately and highly spinning systems. Direct modeling of the joint $(a_1, a_2)$ distribution \citep{2026ApJ...996...71H,2025ApJ...994..261A} independently identifies a high-spin subpopulation at $\sim$15--20\%, broadly comparable to our inferred accretion fraction, in which both components are rapidly spinning, consistent with spin-up by accretion from a circumbinary disk.

Our analysis is largely independent of \citet{2026ApJ...996...71H}, who used GWTC-3 and report that their high-spin subpopulation hinges on GW190517, whereas our sample extends through O4 and is driven primarily by GW231123, with GW190517 one of several supporting candidates. Moreover, the isolated-binary channels (e.g., chemically homogeneous evolution) invoked in \citet{2026ApJ...996...71H} are not expected to produce black holes above the pair-instability mass gap, disfavoring that interpretation as an explanation for the full high-spin subpopulation we identify.

Our results also complement the finding by \citet{VijaykumarFishbach2026} that a hierarchical subpopulation at $q \sim 0.5$ with spins near 0.7 can produce the observed $q$--$\chieff$ correlation \citep{Callister2021}; their inferred hierarchical fraction of $\sim$3\% at low redshift is consistent with our spin-only posterior. The accretion channel contributes high spins across a range of mass ratios and may account for part of the same trend without requiring a specific intrinsic $q$--$\chieff$ correlation. Disentangling the relative contributions of these channels will require joint modeling of masses, spins, and mass ratios.

Several limitations should be noted. Our model fits only spin magnitudes and does not incorporate masses, spin tilts, or redshift into the population likelihood. The secondary spin $a_2$ is weakly constrained for most events and provides limited discriminating power. A joint mass-spin model would enable simultaneous measurement of both the hierarchical and accretion fractions, but requires assumptions about the mass distributions of each channel that are currently poorly constrained. Such a model would also be needed to robustly measure the hierarchical fraction, which our spin-only analysis cannot significantly detect.


\begin{acknowledgments}
The authors would like to thank Christian Adamcewicz, Matthew Mould, Suvodip Mukherjee and Gayathri V. for useful feedback.
Z.H. was supported by NASA grants 80NSSC22K0822 and 80NSSC24K0440. 
This research has made use of data obtained from the Gravitational Wave Open Science Center (GWOSC), a service of the LIGO Scientific Collaboration, the Virgo Collaboration, and KAGRA.
This material is based upon work supported by NSF's LIGO Laboratory which is a major facility fully funded by the National Science Foundation.
\end{acknowledgments}

\bibliography{refs}

\end{document}